\documentclass[aps,prb,showpacs,twocolumn]{revtex4}
\usepackage{graphicx}
\begin{document}

\newcommand{\figureheight}{8.2 cm}
\newcommand{\putfig}[2]{\begin{figure}[h]
        \special{isoscale #1.bmp, \the\hsize \figureheight}
        \vspace{\figureheight}
        \caption{#2}
        \label{fig:#1}
        \end{figure}}

\newcommand{\eqn}[1]{(\ref{#1})}

\newcommand{\be}{\begin{equation}}
\newcommand{\ee}{\end{equation}}
\newcommand{\bea}{\begin{eqnarray}}
\newcommand{\eea}{\end{eqnarray}}
\newcommand{\bean}{\begin{eqnarray*}}
\newcommand{\eean}{\end{eqnarray*}}

\newcommand{\nn}{\nonumber}




\title{ Quenching of Spin Hall Effect in Ballistic nano-junctions 
}
\author{S. Bellucci $^a$ and P. Onorato $^a$ $^b$ \\}
\address{
        $^a$INFN, Laboratori Nazionali di Frascati,
        P.O. Box 13, 00044 Frascati, Italy. \\
        $^b$Department of Physics "A.Volta", University of Pavia, Via Bassi 6, I-27100 Pavia, Italy.}
\date{\today}
\begin{abstract}
We show that a nanometric four-probe ballistic junction can be
used to check the presence of a transverse spin Hall current in a
system with a Spin Orbit coupling not of the Rashba type, but rather due to the in-plane
electric field. Indeed, the spin Hall effect is due to the
presence of an effective small transverse magnetic field
corresponding to the Spin Orbit coupling generated by the
confining potential. The strength of the field and the junction
shape characterize the quenching Hall regime, usually studied by
applying semi-classical approaches. We discuss how a quantum
mechanical relativistic effect, such as the Spin Orbit one, can be observed
in a low energy system and explained by using classical
mechanics techniques.

\end{abstract}

\pacs{72.25.-b, 72.20.My, 73.50.Jt}

\maketitle

\section{Introduction}

The classical Hall effect (HE) is a familiar phenomenon in
condensed matter physics since  E.H. Hall discovered that, when an
electric current flows along a conductor subjected to a
perpendicular magnetic field, the Lorentz force deflects the
charge carriers creating a transverse Hall voltage between the
opposite
edges of the sample. 

Recent developments in the analysis of spin  effects have opened
a new field of research oriented toward the phenomenology of the
so called Spin Hall Effect (SHE).
 In analogy to the conventional
Hall effect, an external electric field may induce a pure
transverse spin current  or result in out-of-plane spin
accumulation near the edges of the sample in the absence of
applied magnetic fields.

This effect has been proposed to occur as a result of the Spin
Orbit (SO) interaction of the electron. In fact the HE arises
physically from a velocity dependent force as  the Lorentz force
while  another velocity dependent force in condensed matter
systems is the SO coupling force\cite{i,ii}. Thus, in
finite-size electron systems  the presence of some kind of SHE can
be due to the interplay between the SO coupling (generating a kind
of Lorentz force)  and the  edge of the device\cite{
noish,qse,iii,iv,v}, analogously to what happens in the Hall
effect.

Several papers discuss velocity dependent forces in
connection to SHE by focusing on their
relativistic quantum
mechanical nature\cite{i,ii}. Here  we start from the SO coupling
which corresponds to the  Hamiltonian \cite{morozb}
\begin{equation}
\hat H_{SO} = -\frac{\hbar}{4 m^* c^2}\;e{\bf E}({\bf r})
\left[\hat{{\bf \sigma}}\times \hat{\bf v}\right]\equiv
-\frac{\lambda^2}{\hbar}\;e{\bf E}({\bf r}) \left[\hat{{\bf
\sigma}}\times \hat{\bf v}\right]. \label{H_SO}
\end{equation}
Here ${\bf E}({\bf r})$ is the electric field, $m^*$  is the effective  electron mass, 
 $\hat{{\sigma}}$ are
the Pauli matrices, the velocity $\hat{\bf v}$ is usually given by
$\left\{\hat{\bf p}-\frac{e}{c}{\bf A}({\bf r})\right\}$, ${\bf
A}$ is the vector potential, ${\bf r}$ is the 3D position vector
and $\lambda^2 = \hbar^2/(2m^* c)^2$.
 The SO interaction has a relativistic nature, because it stems from the
expansion quadratic in $v/c$ of
Dirac equation~\cite{Thankappan} and is due to the Pauli
coupling between electron spin momentum and magnetic
field, which appears in the rest frame of the electron, due to its
motion in the electric field.


Early theoretical studies predicted the  SHE as an {\it extrinsic} effect
due to impurities in the presence of SO coupling \cite{[3]}.  In
this effect SO-dependent scattering off impurities will deflect
spin-$\uparrow$ (spin-$\downarrow$) electrons predominantly to the
right (left).
 More recently, it has been pointed out that
there may exist a different  SHE that, unlike the effect conceived
by Hirsch \cite{Hir},  is purely {\it intrinsic} and does not rely
on anisotropic scattering by impurities. Recently this effect has
been theoretically predicted for semiconductors with SO coupled
band structures as 3D p-doped semiconductors
\cite{[6]} and 2D electron systems with Rashba SO coupling
\cite{[7],cul}.
This SO  contribution, first introduced by Rashba~\cite{Rashba}
and known as $\alpha-$coupling\cite{morozb}, is a natural
coupling which arises due to a structural inversion asymmetry in
quantum heterostructures \cite{Kelly} where 2D electron systems
are realized (2DEG). Experimentally, in $GaAs-AsGaAl$ interface,
values for $ \lambda^2 e E_z $ of order $10^{-11}\; eV\; m$ were
observed \cite{Nitta}. It was shown to be  relevant in low
dimensional semiconductor devices as Quantum Dots (QDs)
\cite{noidot} and Quantum Wires (QWs) \cite{me}.

In  some recent
papers \cite{noish,qse,iii} a different  Spin Orbit coupling term
was investigated ($\beta$-coupling) which arises from the in-plane
electric potential that is applied to pattern a device in  the
2DEG~\cite{Thornton,Kelly}. There, it  was assumed that  the
particle momentum is confined in a 2D geometry, say the $xy$
plane, and the electric field direction is confined in the $xy$
plane as well, with only the $z$ component of the spin entering
the Hamiltonian eq. (\ref{H_SO}). The SO interaction arising from
the lateral confining electric field manifests itself as a weak
 effective magnetic field along the $z$ direction \cite{noish,qse}.

 Because the strength of this
effective magnetic field is quite small, we have to introduce a
device, where relevant effects on the transport properties appear
also at small values of $B_{eff}$. Next we demonstrate that some
observable effects on the spin Hall transport  should be measured
in a nanometric ballistic junction.

The transport through micrometric  ballistic junctions (i.e. a cross
junction between 2 narrow QWs in a 2DEG, also known as a four-probe
junction) was largely investigated about 20 years ago. Several
magneto-transport anomalies were found in these devices, among
these the quenched or negative Hall resistance, bend resistances
and a feature known as the last Hall plateau. The physical origin
of these anomalies may be understood in the B\"uttiker-Landauer
scattering approach \cite{BL} which expresses the resistances in
terms of transmission probabilities.
 These anomalies were largely studied
 and it was shown that many observed effects, e.g. the
classical Hall plateau,  have a classical origin and can be
reproduced, based on classical trajectories.
In the same years, a strong geometry dependence of the transport
properties was shown. In the presence of a transverse magnetic
field the resistances measured in narrow-channel geometries are
mainly determined by the scattering processes at the junctions
with the side probes which depend strongly on the junction shape
\cite{ref289}. This dependence of the low-field Hall resistance
was demonstrated \cite{ref358} and measured \cite{fordbvh}.

\

Some papers in recent years calculated the spin-Hall conductance
in a 2D junction system with Rashba SO coupling and disorder,
using the four-terminal Landauer-B\"uttiker
formula\cite{shm1,shm2,shm3}. In this paper we discuss the effects
of the non-Rashba $\beta-$SO term in the absence of disorder in a
2D crossed nanojunction. In  section II we introduce the model of
the confining potential, linked to the reliable devices,  and the
corresponding  effective field. In section III we discuss the
quenching of the Hall effect obtained for our model and thus the
extended calculation to the quenching of the Spin Hall effect.

\section{Effective Field and Confining Potential }


\subsection{Effective Field}

When we take in account the $\beta-$SO coupling  an electric field
$\bf E(\bf r)$ was obtained starting from the transverse potential
confining electrons to the 2D device, $V_c$ (due to the split gate
electrodes) as ${e\bf E}({\bf r})={\bf\nabla} V_c({\bf r}) $. Next
we  assume $\langle p_z\rangle=0$ and neglect $E_z$ i.e. the
$\alpha-$coupling. Hence, the general Hamiltonian of a single not
interacting electron has the form \bea H&=&\frac{\bf{p}^2}{2 m^*
}+\frac{\lambda^2}{\hbar}e\left({\bf
E}(\bf r)\wedge {\bf p}\right)_z \sigma_z+V_c({\bf r})\nonumber \\
&=& \frac{{\bf \pi}^2}{2}
+V_c({\bf r}) -\frac{\lambda^4 m^*}{2 \hbar^2}{\left|{\bf
E}\right|^2}, \eea where $$
\pi_i=(p_i-\epsilon_{ijz}\frac{\lambda^2}{\hbar}m^* e E_j\sigma_z)
.
$$  
 The commutation
relation,  \bea \label{ecom} \left[\pi_x,\pi_y\right]=-i \hbar
\left(\frac{\lambda^2}{\hbar}m^* e {\bf \nabla}\cdot{\bf E}\right)
\sigma_z \equiv  -i \hbar \frac{e}{c}{B}_{eff}\sigma_z \eea is
exactly equivalent to the usual commutation rule  of a charged
particle in a  transverse magnetic field, where the two different
spin directions experience the opposite directions of the {\em
effective}  field ${B}_{eff}$.

\subsection{Confining potential and strength of the effective field}

In a cross junction sample, the confining electrostatic potential
$V_c$ for an electron is not exactly known. However, it is
plausible that there has to be a potential minimum at the center
of the junction. In this respect, it would be appropriate to
qualitatively model the smooth potential walls as
 \bea V_c(x,y)=
\frac{m^*}{2}\omega_d^2 R^2 \frac{x^2 y^2}{(R^2+x^2)(R^2+y^2)},
\eea where $\ell=\sqrt{\frac{\hbar}{m^* \omega_d}}$ can be related
to the effective width of the wires, $W$ and $R$ to the effective
radius of the crossing zone. It is known that the effective width
$W$ corresponds to a real width $W_R$ some times larger than $W$
and can be further reduced by acting on the split gate electrodes.

In general we can relate the frequency, $\omega_d$ to $W$ as \bea
\omega_d \sim \frac{(2\pi)^2}{2}\frac{\hbar}{m^*
W^2},\label{omegad}\eea obtained by a comparison between the
energy levels of a harmonic oscillator and a square potential
well. Notice that the measurements about energy levels in
mesoscopic devices confirm the fact that the effective width  is
smaller than the real size of the device, e.g. in the QD
discussed in \cite{taru} the real diameter is $D\sim 500 nm$ while
$\hbar \omega_d$ corresponds to $W\sim 100 nm$.
{From eq.(\ref{ecom})  it follows that the strength of the
effective magnetic field can be obtained as
\begin{equation}
{B}_{eff}\sim \frac{(2\pi)^4}{4}\frac{\hbar c}{e}
\frac{\lambda^2}{W^4}
\end{equation}
 For narrow wires  of lithographical width
ranging from $20$\cite{kunze} to $200\, nm$ patterned in the usual
semiconductor heterostructures  it is possible to obtain values of
$B_{eff}$ corresponding to $\omega_{eff}/\omega_d\sim 10^{-7}-
10^{-3}$, where $\omega_{eff}=e B_{eff}/(m^* c)$. In fact we can
estimate the effective value of $\lambda$ in the 2DEG starting
from the measured value of $\alpha$ in
literature\cite{shapers,HgTe}. For GaAs heterostructures
$\lambda^2$ is $\sim 10^{2}nm^2$, one order of magnitude less than
in InGaAs/InP heterostructures where $\lambda^2$ takes values
between $0.5$ and $1.5\,nm^2$ (in agreement with the values used
in ref.\cite{morozb} where   $W\sim 200 nm$) while  for HgTe based
heterostructures  $\lambda^2$ can be one or more order larger up
to some tens of  nm$^2$\cite{Schultz,HgTe}. }

However, it is clear that this effect could be larger than the
Rashba effect in some appropriate samples.

\begin{figure}
\includegraphics*[width=.85\linewidth]{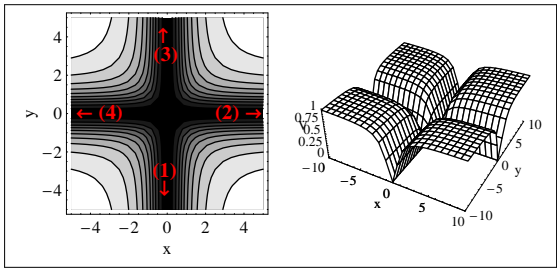}
 \caption {(Color online) Contour and 3D Plot of the potential  $V_{c}(x,y)$. }
\end{figure}

\section{Calculations and results}

In this section we want to investigate the effects of a quite
small effective magnetic field, so we have to analyze in more
details the so called quenched region. It corresponds to  the
regime where, in the presence of a quite small  magnetic field,
$B$, the "quenching of the Hall effect" was measured, i.e. a
suppression of the Hall resistance  or "a negative Hall
resistance" \cite{fordbvh}, $R_H$.

In order to pursue our aim we first discuss the known case of the
quenching of the Hall effect. A comparison with theoretical and
experimental results carried out in the past allow ourselves to
test our approach.

\subsection{Quenching of the Hall Effect}

\begin{figure}
\includegraphics*[width=.75\linewidth]{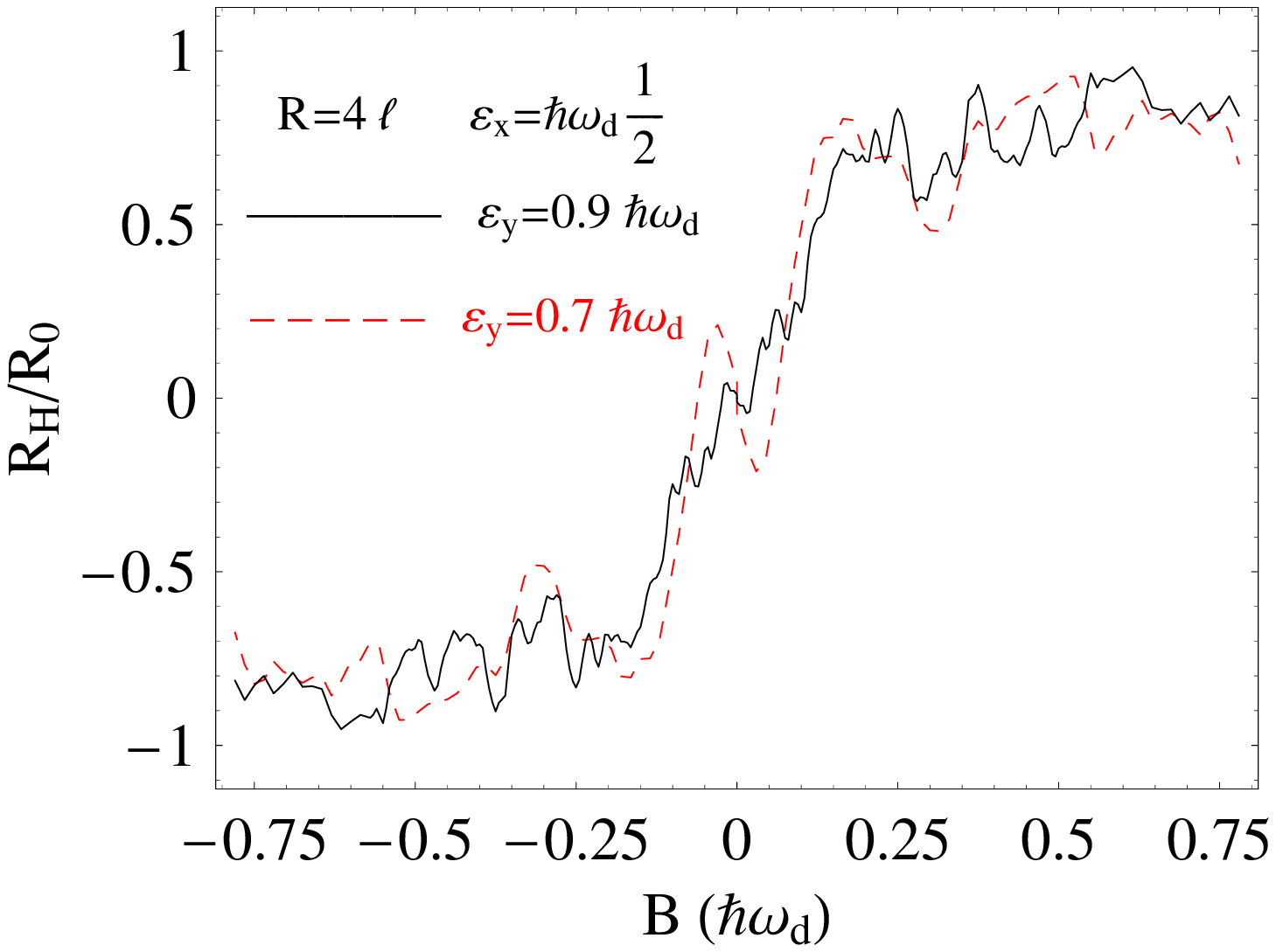}
\includegraphics*[width=.75\linewidth]{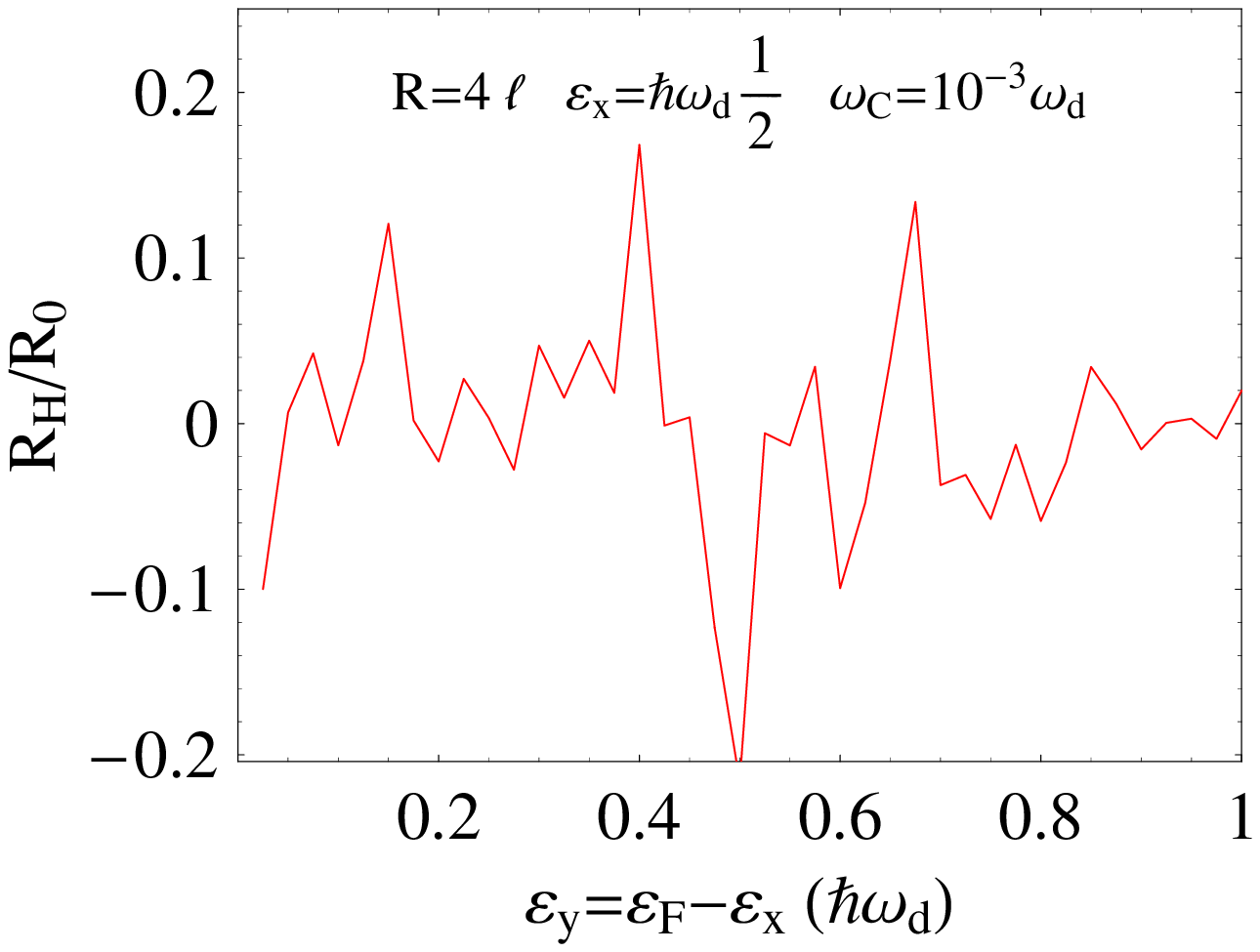}
 \caption {(Color online) (Top) $R_H\,vs\, B $  for two different values of the Fermi energy (experimentally corresponding to the gate Voltage, $V_g$).
Here $B$ corresponds to  $\hbar \omega_c=\hbar e B/(m^* c)$ in
units $\hbar \omega_d$.  At small $B$ it is clear the presence of
a quenched region where the Hall resistance, $R_H$, is negative.
For the device measured in ref.\cite{fordbvh}
  the crossing wires have  a real width  $W_R\sim 200 nm$
   and an effective one $W\lesssim 60 nm$.
   It follows that $\hbar \omega_d\sim 6 meV$ while the range of $B$
    ($\pm .75 \hbar \omega_d$ ) corresponds to a $B$ value which ranges between
    $-1$ T and $1$ T (see \cite{noidot}).
    This allows for a comparison with the results shown in Fig.1 of
    ref.\cite{fordbvh}.
 (Bottom) $R_H$ as a function of the Fermi energy at a value of magnetic field consistent with the effective magnetic field predicted.
 The irregular oscillations are in agreement with the calculation in ref.\cite{ref358} . The value of $y_0$ is fixed at $\sim -10 \ell$.}
\end{figure}
In a four-fold symmetric junction, as the one shown in Fig.(1),
Hall resistances follow from the B\"{u}ttiker formula
\cite{BL} as a function of the transmission probabilities $T_{ij}$
across the junction from the reservoir $i$ to lead $j$ as
\cite{gei92} \bea \label{RI}R_H= R_0
\frac{T_{21}-T_{41}}{T_{21}^2+T_{41}^2+2
T_{31}(T_{21}+T_{31}+T_{41})}, \eea with $R_0\propto h/e^2$.
Our calculations are based on a simulation of the classical
trajectories of a large number  of electrons with the Fermi
energy, $\varepsilon_F$, to determine the classical transmission
probabilities, $T_{ij}$. These coefficients can be determined from
classical dynamics of electrons injected in lead $1$ i.e. at
$y_0\ll-R$ where $V_c(x,y_0)\sim m^* \omega_d^2 x^2/2$. Next,
we restrict ourselves to one channel, i.e. the lowest transverse mode.
Thus, we choose an injection probability as
$$
p_0(x_0,y_0)\propto e^{-\frac{x_0^2}{\ell^2}}
$$
corresponding to the asymptotic  eigenstate of the single electron
in the potential $V_c$ without  magnetic field. It follows that
the energy can be written as $\varepsilon_F=\varepsilon_x+
\varepsilon_y$ ($\varepsilon_x\approx \hbar \omega_d/2$) and
$$
m^*{\bf v}_0\approx\left(\pm \sqrt{{2 m^*\varepsilon_x-{m^*}^2
\omega_d^2x^2}};\sqrt{{2 m^*\varepsilon_y}}-m^*\omega_c x\right).
$$
Thus, we have calculated $T_{ij}$ determined by numerical
simulations of the classical trajectories injected into the
junction potential $V_c$  with boundary conditions ${\bf
r}(0)\equiv (x_0,y_0)\;;{\bf v}(0)\equiv {\bf v}_0$, each one with
a weight $p_0(x_0)$.

$R_H$ is reported in Fig.(2) as a function of an external magnetic
field, $B$. The presence of the quenching region near $B=0$ is
shown for two different values of the Fermi energy. In this case
the calculation does not take into account the SO interaction.

Our results can be compared with the experimental data reported in
ref.\cite{fordbvh} and support the validity of our approach.

\subsection{Quenching of the Spin Hall Effect}

Next, we can calculate what happens in a vanishing external
magnetic field, just taking into account the SO effect. This is
shown in Fig.(3.bottom). In this case the effective magnetic field
is fixed by the geometry and the nature of the sample, and we just
can change the Fermi energy. Results are reported just for the
lowest transverse mode.

\begin{figure}
\includegraphics*[width=.505\linewidth]{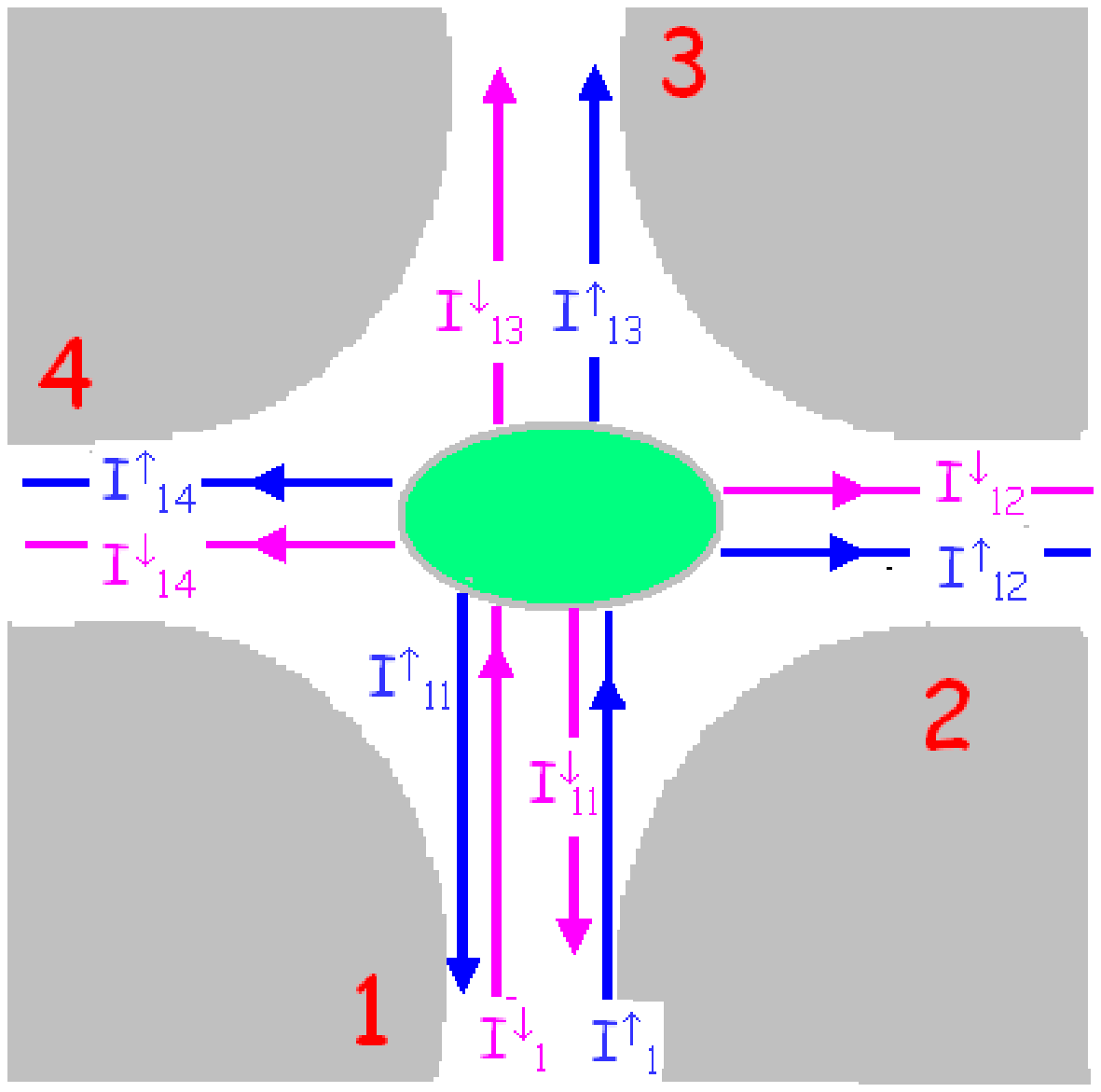}
\includegraphics*[width=.75\linewidth]{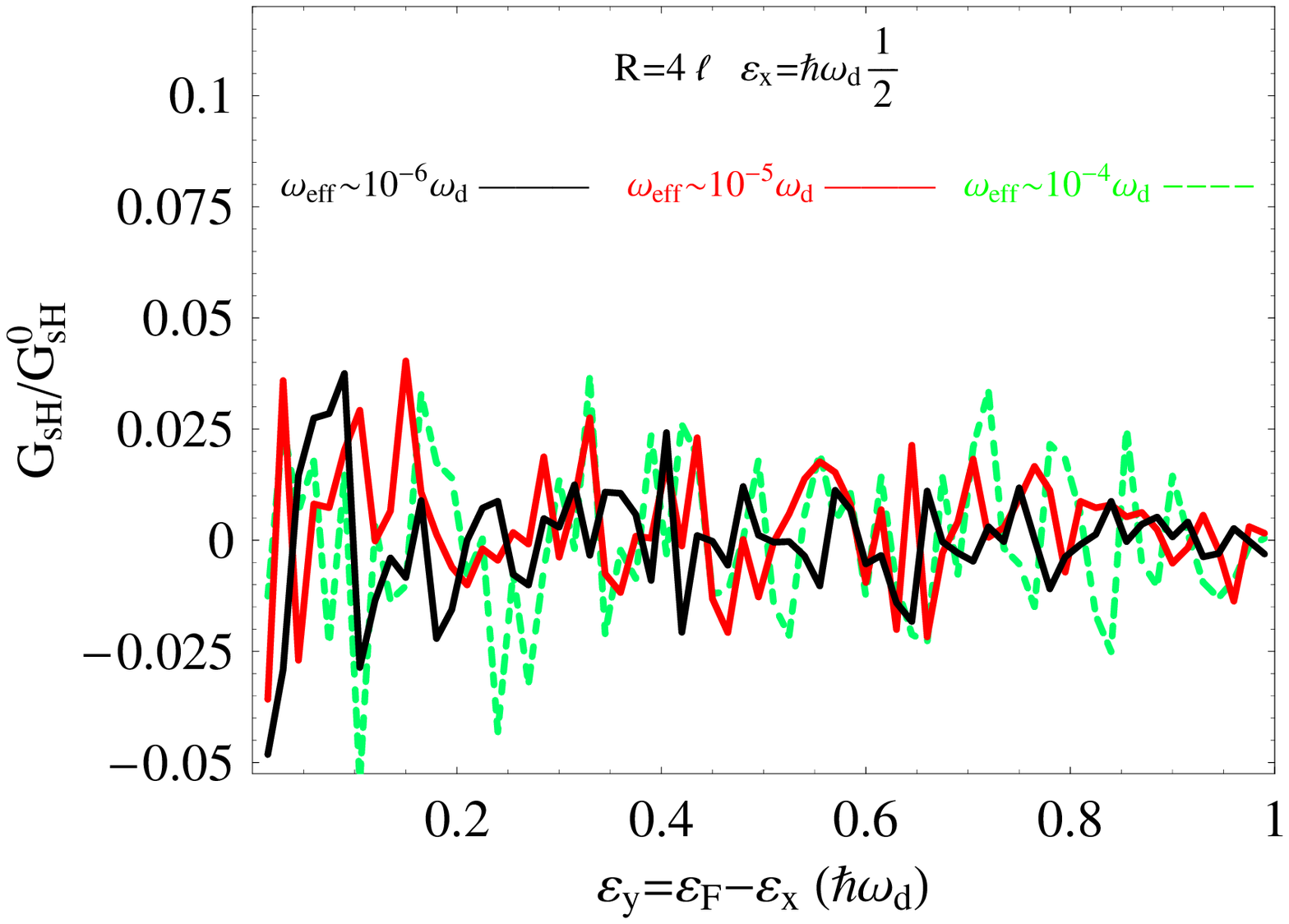}
 \caption {(Color online) (Top) Schematic representation of spin currents in
 the four-probe
junction.
 (Bottom) Spin Hall conductance, $G_{sH}$, as a function of the Fermi energy under an
  inhomogeneous effective
  magnetic field consistent with the assumed sample.
 The irregular oscillations are in agreement with Fig.(2.b) and calculations in ref.\cite{ref358}.
 The non uniform magnetic field behaves quite similarly to the uniform one. The value of $y_0$ is fixed $\sim -10 \ell$.
 Note that oscillations are not much suppressed by the reduction of the
 ratio $\omega_{eff}/\omega_d$ (here $\omega_{eff}\equiv eB_{eff}/m^* c$).}
\end{figure}

Now we can discuss in detail the currents in the sample. The
current $I_i$ in the lead $i$ of a four-probe junction with
chemical potentials $\mu_j=eV_j$ attached to leads $j$ can be
expressed in terms of the $T_{ij}$ by $I_i = e^2/h\sum_j
T_{ij}(V_i-V_j)$, and normalization requires $\sum_j T_{ij}=1$
\cite{BL,gei92}. We follow the schematic representation in
Fig.(3.top), where the currents corresponding to opposite spin
polarizations are located at the opposite edges of each probe,
according to the edge localization discussed in ref.\cite{noish}.
 $I_1=I_1^{\uparrow}+I_1^{\downarrow}$ is the
injected current, which is localized on the right-hand side of the
wire $1$, $I^s_{i1}$ is the charge current outgoing from  the lead
$i$ corresponding to the spin polarization $s$. Thus, there should
be two spin polarized (charge) currents, $I_H^s$ in the $x$
direction, from right to left, given by $I_H^s=I_{41}^s-I_{21}^s$.
When we take into account a spin unpolarized injected current,
$I_1$, it follows from the spin dependence of the effective
magnetic field that
$I_{11}^{\uparrow}=I_{11}^{\downarrow}$,$I_{31}^{\uparrow}=I_{31}^{\downarrow}$,
$I_{21}^{\uparrow}=-I_{21}^{\downarrow}$ and
$I_{41}^{\uparrow}=-I_{41}^{\downarrow}$. The symmetry of the
device implies that the charge Hall current vanishes,
$I_H=I_H^{\uparrow}+I_H^{\downarrow}=0$. In this case we can
define also the Spin Hall current as   \bea\label{ish}
I_{sH}=\frac{\hbar}{2e}\left(I_{H}^{\uparrow}-I_H^{\downarrow}\right).
\eea
 This result can be also obtained by calculating the response
of the spin current operator \cite{newref} $$\hat{\bf
J_s}=\frac{\hbar}{4}(\hat{\sigma_z}\hat{\bf v} + \hat{\bf
v}\hat{\sigma_z})$$  to the electric field. This calculation can
easily be done within the framework of the Landauer formalism and
give the current in eq. (\ref{ish}). Although this may not be  true
in the general case, where $\hat{\bf v}$ does not commute with
$\hat{\sigma_z}$, nonetheless it holds valid in our case.
Thus  a spin current, linked to a vanishing charge current, is now
present in the four-probe junction. Starting from eq. (\ref{ish})
we can also define the corresponding spin Hall resistance by using
eq. (\ref{RI}) where $ R_{sH}^0=\frac{2e }{\hbar}R_0 $ appears
instead of $R_0$
 \bea \label{Rs}
 R_{sH}= R_{sH}^0
\frac{1}{T_{21}^\uparrow-T_{41}^\uparrow}. \eea The latter
formula is in agreement with the results obtained in ref\cite{ii}
(see eqs.(10c) and (11c)). The results of our calculation reported
in Fig.(3.bottom) show
 oscillations of $G_{sH}=1/R_{sH}$ with quite nonvanishing values. From the figure
 it is manifest that the oscillations are not much suppressed from the reduction of the
 ratio $\omega_{eff}/\omega_d$.

 \section{Conclusions}

We wish to add here some useful remarks on rather relevant issues.
The four-probe junction system appears to be like a kind of
ultra-sensitive scale, capable of reacting to the smallest
variations of the magnetic field. In this case, any breakdown of
the symmetry (left right 12-14) produces a Hall current. Since we
are in the quenching regime, such a current cannot be predicted,
either in its intensity, nor in its orientation. For the above
mentioned reasons, the effect is observable even for what concerns
the SO effect yielding effective fields which are very small (with
respect to $\omega_d$).
 Further calculations will allow us to discuss the effect
 of the higher transverse mode and those due to different
 geometries as discussed in ref.\cite{fordbvh}. However the
 results obtained here should be experimentally confirmed by
 giving the signature of an intrinsic SHE due to the
 in plane electric field.

In conclusion, in this paper we considered the SHE in a small
ballistic device, in which the SO coupling arises due to the
in-plane confining potential, in contrast to the more widely
studied situation of the Rashba SO interaction. We believe that
this problem is worth studying, especially in the context of how
the strong electric fields near the edges of the confining
potential may affect the spin accumulation due to bulk spin
currents. Our main result is summarized in the bottom panel of
Fig.3 above. There is a spin Hall resistance, which oscillates as
a function of the Fermi energy. As we pointed out in the above,
the magnitude of these oscillations does not seem to depend on the
strength of the effective SO field. This interesting fact may be
connected to the discussion of electron trajectories in the
quenching Hall regime carried out for instance in
\cite{fordbvh,gei92}. However we do not address this
correspondence in detail. As well known, SHEs can  be observed by
the spin accumulation at the boundaries they produce. It is also
well known, though, that the pure spin Hall current flowing out
between the transverse probes is surely connected to a spin
accumulation in them. The value of $\langle S_z (x) \rangle $
which we found in the probes is of order 10$^{-2}\hbar$/2 for
$\omega_{eff}/\omega_d$ values reported in Fig.(3.bottom) as we
show in Fig.4.

\begin{figure}
\includegraphics*[width=.9505\linewidth]{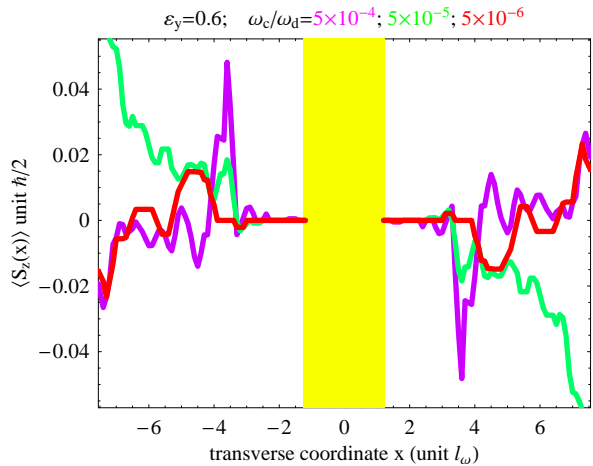}
 \caption { (Color online) The
one-dimensional transverse spatial profile of the spin
accumulation $\langle S_z(x) \rangle$ across the transverse probes
(2 and 4) of the ballistic junction with vanishing
$\alpha$-coupling while the $\beta$-SO coupling corresponds to
$\omega_c/\omega_d$ ranging from $10^{-6}$ to $10^{-4}$.}
\end{figure}


\bibliographystyle{prsty} 

\bibliography{}

\end{document}